\documentclass{article}
\usepackage{geometry}
\usepackage{multirow}
\usepackage{graphicx}
\usepackage{epstopdf}
\usepackage{amsmath}
\usepackage{lscape} 
\usepackage{apacite}
\usepackage{natbib}
\usepackage{xcolor}
\usepackage{cite}
\usepackage{amssymb}
\usepackage{indentfirst}
\usepackage{tikz}
\usepackage{pdfpages} 
\usepackage[colorlinks,
linkcolor = blue,		
anchorcolor =blue,	
urlcolor = blue,		
citecolor =blue,		
]{hyperref}
\usepackage{fancyhdr}
\usepackage[colorlinks, citecolor=blue]{hyperref}
\usepackage{float}
\title{\textbf{Research on the Impact of Innovative City  and Smart City Construction on Digital Economy:Evidence from China}}
\author{ Zhanpeng Huang }
\date{}

\usepackage{booktabs}
\usepackage{array, caption, threeparttable}
\usepackage[font=small,labelfont=bf,labelsep=none]{caption}

\usepackage{fancyhdr}
\usepackage{pdflscape}
\usepackage[absolute]{textpos}
\makeatletter
\fancypagestyle{landscape}{
    \fancyhf{}

    \setlength{\TPHorizModule}{\paperwidth}%
    \setlength{\TPVertModule}{\paperheight}%
}
\makeatother

\begin{document}
	\maketitle
	
	\begin{abstract}
		
		Does the national innovation city and smart city pilot policy, as an important institutional design to promote the transformation of old and new dynamics, have an important impact on the digital economy? What are the intrinsic mechanisms? Based on the theoretical analysis of whether smart city and national innovation city policies promote urban digital economy, this paper constructs a multi-temporal double difference model based on a quasi-natural experiment with urban dual pilot policies and systematically investigates the impact of dual pilot policies on the development of digital economy. It is found that both smart cities and national innovation cities can promote the development of digital economy, while there is a synergistic effect between the policies. The mechanism test shows that the smart city construction and national innovation city construction mainly affect the digital economy through talent agglomeration effect, technology agglomeration effect and financial agglomeration effect.
		
		\begin{center}
		\end{center}
		
	\end{abstract}	
	
	\section{Introduction}	
	
	With the development of Internet technology, the integration of information technology and industry continues to accelerate, and the research on digital economy has gradually increased in recent years. Some studies point out that the digital economy is an important indicator to measure people's welfare level, and should focus on the measurement of digital consumption\citep{Strassner2020Measuring}
At the same time, the digital economy is still under debate in terms of standards and definitions. In terms of types of services, the digital economy encompasses a wide range of concepts, such as e-commerce, e-government, e-payment systems, e-banking, e-knowledge processing, internet banking, mobile banking, etc.\citep{mesenbourg2001measuring}.
	These businesses are based on industrial IoT and consumer Internet technologies, showing the integrated development of manufacturing and service industries. This trend has become more evident after the pandemic\citep{Jiang2020Digital};
	It can be seen that the combination of technology and industry is one of the characteristics of the digital economy. And in the long term, this feature is manifested in complex dynamic changes, which is why the digital economy is difficult to measure under the long term.\citep{Carlsson2004Digital}.
Nonetheless, no matter how the industries associated with the digital economy change, it all boils down to a combination with Internet technology, and we can measure it by quantifying the Internet element when we measure it.
	
	Based on the theoretical analysis of whether smart city and national innovation city policies promote urban digital economy, this paper constructs a multi-temporal dual difference model with a quasi-natural experiment using urban dual pilot policies and systematically investigates the impact of dual pilot policies on the development of digital economy. It is found that both smart cities and national innovation cities can promote the development of digital economy, while there are synergistic effects between the policies. The mechanism test shows that the smart city construction and national innovation city construction mainly affect the digital economy through talent agglomeration effect, technology agglomeration effect and financial agglomeration effect.
	
	The remainder of the paper is structured as follows: Part II is the policy background and mechanism analysis; Part III is the empirical study; and Part IV is the conclusion and policy implications. The charts not detailed in the paper are attached in the appendix at the end of the paper.
	\section{Literature Review }
	
	The two pilot policies, National Innovative City Pilot and Smart City, are both important strategies for China's urban development.Smart cities represent a conceptual urban development model based on leveraging human, collective and technological capital for the development and prosperity of urban agglomerations\citep{Angelidou2014Smart}.
	With the increasing number of cities included in the scope of pilot city construction and the increasing time of policy implementation, scholars have gradually started to evaluate the effects of this policy. At present, domestic studies mainly focus on innovation capacity and sustainability assessment, mainly using spatial measures and double difference methods.
	European smart city policies drive local companies, especially multinationals, to translate technology into local needs, thus promoting higher levels of innovation\citep{Caragliu2019Smart}.From the perspective of spatial synergy, based on the triple helix theory of industry-academia research and the theory of spatial association, the infrastructure of China's innovative cities also affects economic development, which reveals the source of differences in the effects of policies between cities and the logic of the inner policies at work in cities\citep{Wang2021Collaborative}.The innovative city pilot policy also has a positive effect on urban green innovation and has an "inverted U-shaped" shape\citep{Li2022Does}.The National Innovation Pilot Policy (NICP) has a long-term positive impact on innovation performance in local and neighboring regions, while the spillover effect exists only in provincial cities and geographically adjacent cities in neighboring provinces, in line with the law of distance decay\citep{Gao2022Government}.
	
	In summary, although studies have found that pilot city policies have an important impact on urban economic development and also include assessments of ecological protection, innovation capacity, and spatial spillover, existing studies have not yet paid sufficient attention to the role of pilot policies on the development of the digital economy, and less attention has been paid to the synergy between the two policies. In fact, smart cities and innovation cities complement each other in different dimensions and have strong synergistic effects.
	
	\section{Analysis of policy background and its mechanism}
	\subsection{Policy Background}
	After the reform and opening up, China urgently needs to rely on scientific and technological progress and innovation to achieve social transformation. In 2005, the State Council mentioned for the first time in the document \textit{Outline of National Medium and Long-Term Science and Technology Development Plan (2006-2020)} (Guo Fa [2005] No. 44) that the ability of independent innovation should be placed in a prominent position in all science and technology work. The document clearly establishes the technology development mode of establishing circular economy in key cities, improves urban traffic management system, enhances the level of integrated urban management, and develops urban digital integration management technology.
	
	In 2010, \textit{the Ministry of Science and Technology promulgated the Guidance on Further Promoting Innovative Cities} (Guo Ke Fa Ti [2010] No. 155), which pointed out that cities are the center of regional economic and social development, the gathering place of various innovative elements and resources, and their development has a great impact on the overall development of the region and the country. The development of cities has a great impact on the overall development of the region and the country, and a number of cities should be promoted to take the lead in the ranks of innovative cities to form a model for more cities to embark on the path of innovative development. By 2020, the two ministries will have approved 78 national innovative pilot cities (districts), including 72 prefecture-level cities, 4 urban areas of municipalities directly under the central government and 2 county-level cities.
	
	Under the influence of the global financial crisis in 2008, IBM first put forward the new concept of smart earth and as a smart project has been taken by countries around the world as a key area to cope with the international financial crisis and revitalize the economy. In 2009, Premier Wen Jiabao treated \textit{"Internet of Things" as an important new industry in "Let Science and Technology Lead China's Sustainable Development"} and established logistics technology R\&D centers in Wuxi, Shanghai, Nanjing, Wuhan, Shenzhen and Chongqing respectively. According to the \textit{National New Urbanization Plan (2014-2020)} and \textit{the Guidance on Promoting the Healthy Development of Smart Cities} (Fa Gai Gao Ji [2014] No. 1770) jointly issued by the Development and Reform Commission and other 8 departments, there will be 290 smart cities by 2020, which will be released in three batches.
		
		\begin{figure}[h]
		\begin{center}
			\includegraphics[width=0.7\textwidth]{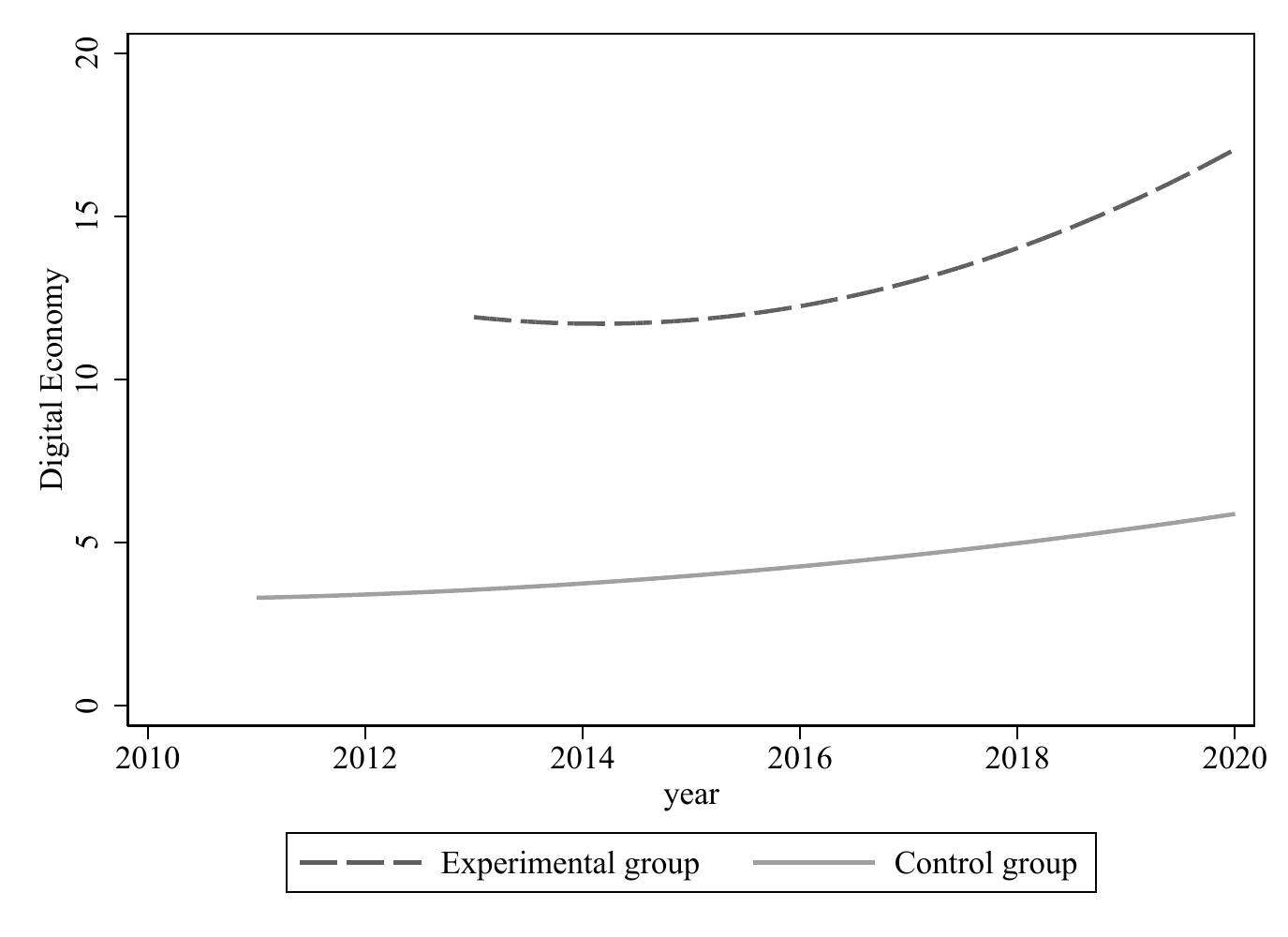}
			\caption{Comparison chart of dual pilot and non-dual pilot} 
			\footnotesize Note: The reason for the inconsistent starting point of the data for the experimental control group is that the data for the experimental group of smart cities and national innovative simultaneous pilot cities started from 2013 at the earliest.
		\end{center}
		\end{figure}
	
	Both innovative cities and smart cities need Internet technology as support, and the Internet is one of the core of the digital economy. Innovative city pilot promotes entrepreneurial vitality and industrial structure optimization, coupled with smart cities to enhance the logistics system well to strengthen the use of data, ultimately forming a synergy in the development of the digital economy.
	
	Comparing the level of digital economy in cities with simultaneous pilots and cities with non-simultaneous pilots, it can be seen from Figure 1 that the level of digital economy in the pilot cities (experimental group) is more developed relative to the non-pilot cities (control group) after the dual pilots.

	\subsection{Mechanistic Analysis}
	
	The mechanism of national innovation city and smart city pilot policies affecting urban digital economy, as shown in Figure 2, is mainly reflected in the following three aspects:
	
	First, the development of new economic sectors is limited by the quantity and structure of financial investment. The pilot policies of smart cities and innovative cities can bring good expectations to financial capital and generate financial agglomeration effect. Traditional financing channels tend to be more fragmented due to information asymmetry and risk prevention considerations, while the pilot policy itself is a good development signal to support innovation and entrepreneurship on the one hand, and the development of investment institutions on the other. At the same time, the pilot policy can alleviate the demand of Internet enterprises through the financial agglomeration effect, which in turn promotes the development of the digital economy. According to the study, it can be found that the internal network structure formed by urban financial agglomeration can benefit the financial industry and related industries at the same time. For example, Shenzhen has established the Shenzhen Venture Capital Guidance Fund in the Notice of the Shenzhen Municipal People's Government on the \textit{Issuance of the Implementation Plan of Shenzhen National Innovative City Master Plan}, Shenzhen Government [2009] No. 9. This implies that the pilot city has good financial absorption capacity.

	Second, talent clustering. The structure of urban real estate investment has a gating effect on the concentration of scientific and technological talents, and the planning of cities plays an important role in the concentration of talents and scientific and technological innovation. Smart cities enhance the comfort of residents by integrating the city's component systems and services. With the new Internet and the Internet of Things, smart cities can inform, industrialize and urbanize. Smart city pilots facilitate improved investment in urban real estate, which in turn affects the concentration of talent. Based on the traditional push-pull theory, population mobility is in pursuit of a better standard of living. In 2022 the national evaluation index of smart cities released mentions transportation services and information networks. The improvement of urban living conditions will greatly enhance the attractiveness of talent inflow. Lianyungang's "Several Policies on Deepening the Construction of Innovative Cities" emphasizes the extensive gathering of innovative and entrepreneurial talents, creating a superior environment for their development, attracting talents through subsidies and incentives in various ways and implementing various special strategies. The preferential policies for talents in the pilot innovative cities have lowered the entry barrier for talents. Smart cities and innovative cities form a synergy that both lowers the entry barrier for talent and enhances the attractiveness of the city to talent.

	Third, technology concentration. In 2010, the Ministry of Science and Technology released the "Innovative City Construction Monitoring and Evaluation Index System (for trial implementation)", which includes "the proportion of local financial allocation for science and technology to local financial expenditure" and "the number of granted invention patents per million people" as acceptance indexes. This has provided an incentive for the pilot cities to produce technological innovation results, and promoted the clustering of regional technological innovation results. The pilot policy also emphasizes strengthening the protection of intellectual property incentives. The perfect property rights protection system and incentive system in the pilot cities are conducive to technology agglomeration.
	
		\begin{figure}[h]
		\begin{center}
			\includegraphics[width=0.8\textwidth]{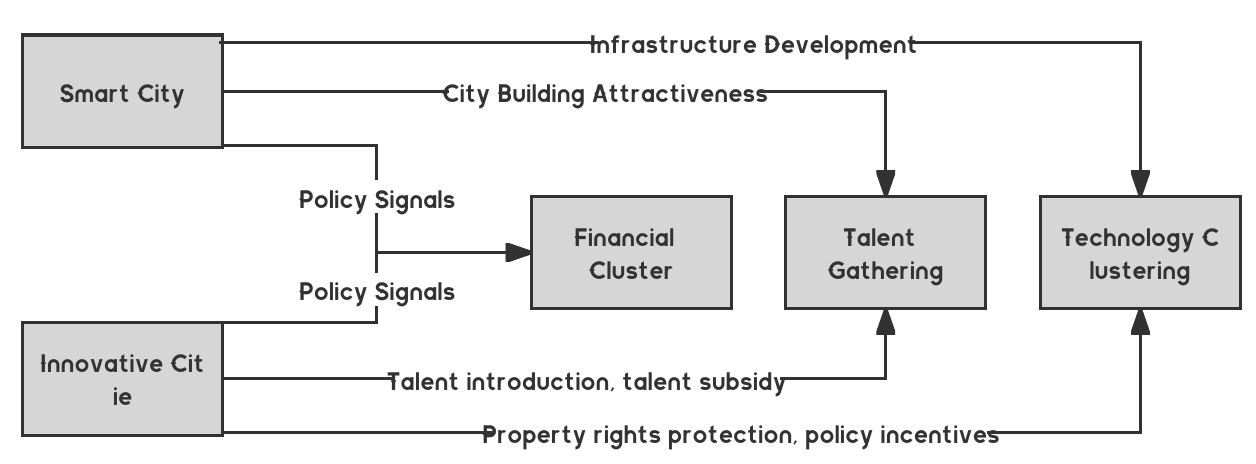}
			\caption{Impact Path} 
		\end{center}
	\end{figure}

	\section{Estimation strategy, variable settings and data sources}
	
	\subsection{Multi-temporal did baseline regression model}
	
	The National Innovative Cities Pilot Policy began in 2008 with a gradual expansion approach to approval, and now has 78 cities. The Smart City policy started in 2012 and three batches were announced between 2013 and 2014, with a total of over 290 cities. Both city pilot policies are exogenous policy shocks to the development of the digital economy, while having synergistic effects. In this paper, we adopt a multi-period double difference method to estimate the impact of the dual pilots of smart cities and national innovation cities on the level of urban digital economy development, and construct the following regression model:
	
	\begin{equation}\label{1}		 {DEI}_{i,t}=\alpha+\beta{Wtreat}_{i,t}+\gamma{Control\_Var}_{i,t}+CityFE+YearFE+\varepsilon_{i,t} 
	\end{equation}
	
$DEI$ indicates the level of development of the digital economy, ${Wtreat}_{i,t}$ indicates the National Innovation City and Smart City dual pilot feature value. $Control\_Var$ is the ensemble of control variables.CityFE is the city fixed effect and YearFE is the year fixed effect. $\varepsilon$ is the random perturbation term.Estimated coefficient $\beta$ Reflects the average difference in the level of digital economy development of cities before and after the impact of the implementation of the dual pilot policy for smart cities and national innovation cities.		

	\subsection{Variable Settings}
	\subsubsection{Explained variables}
There are still no uniform standards for measuring the size of the digital economy. Most articles measure data at the provincial and national levels, and there are still relatively few measurements at the municipal level.The EU measures tier 1 indicators in five main areas such as broadband access, human capital, Internet adoption, digital technology adoption and the extent of digital public services in each country\footnote{From the 
	\textit{Digital Economy \& Society in the EU}, published in 2014, which refers to the \textit{Digital Economy and Society Index} (DESI)}
The U.S. Department of Commerce Digital Economy Advisory Board's 2016 (DEBA) \textit{First Report of the Digital Economy Commission} proposes to measure the impact of digitization on economic indicators  (GDP, productivity levels, etc.), and the role of digitization in scaling across industries.
Comparing the international indicators, Europe has a more detailed data acquisition, while the US has a richer measurement perspective. China, on the other hand, is currently more specific in the scope of urban digital economy measurements, favoring industries and cities.	

In this paper, the Internet is taken as a key measurement element, and four indicators, namely Internet penetration, related employees, related output and cell phone penetration, are used, drawing on the method of ~\citet{Zhao2020Shuzi}.Combining the availability of relevant data at the city level, we measured the comprehensive development level of digital economy in terms of Internet development and digital financial inclusion, and finally constructed the following indicators: the number of cell phone users, the number of Internet users per 100 people, telecommunication business income, digital financial inclusion index, the number of relevant employees, and the number of granted invention patents. The original data of the above indicators come from the China City Statistical Yearbook. The digital potential refers to the China Digital Inclusive Finance Index, which is jointly compiled by the Digital Finance Research Center of Peking University and Ant Financial Services Group ~\citep{Guo2020Cedu}. 
The comprehensive digital economy development index, denoted as DEI, is obtained by standardizing and then downscaling the data of the above six indicators through the method of entropy weight method analysis.

	\begin{table}[h]
	
	\caption{ Digital Economy Indicators}
\begin{center}
\begin{center}
			\begin{tabular}{ccccc}
		\hline
		Tier 1 & Tier 2 & Tier 3  & Properties & Weights \\ \hline
		Basic &  Phone Penetration &  phone subscribers& + & 0.091 \\ 
		
		~ & Internet consumers &  Internet users & + & 0.101 \\ 
		
		 Output & Internet-related outputs & business revenue & + & 0.138 \\ 
		
		~ &  related outputs & Digital Finance Index & + & 0.031 \\ 
		
		Development & Talent Support & related practitioners & + & 0.292 \\ 
		
		~ & Technical Support &  Invention Patents Granted & + & 0.342 \\ 
		
		\hline
	\end{tabular}
\end{center}
\end{center}
\end{table}

\subsubsection{Core explanatory variables}

In this paper, the dual pilot of smart city and national innovation city  ${Wtreat}_{i,t}$  are the core explanatory variable.When the city is set as a national innovation city pilot in the same year, the variable  ${Innovation\_treat}_{i,t}$ is assigned to 1 for the same year and later, and 0 for the rest.Similarly, when a city becomes a smart city pilot, the corresponding variable ${Smart_treat}_{i,t}$ is assigned a value of 1 and the rest is assigned a value of 0.The interaction term between the innovation city pilot dummy variable and the smart city pilot variable (${Innovation\_treat}_{i,t}\times{Smart_treat}_{i,t}$) is used to characterize the policy treatment effect (${Wtreat}_{i,t}$) of the dual pilot. When the city is a dual pilot of both smart city and national innovation city, ${Wtreat}_{i,t}$ is 1 and the rest is 0.

\subsubsection{Control variables}

Considering that other city characteristic factors may have an impact on the level of digital economy development, this paper controls for the following variables:

$ 1. $ Level of economic development. Developed economic regions facilitate technology clustering and talent pooling, and promote the development of digital economy. In this paper, we use the logarithmic $gdp$ per capital to measure the level of economic development ($GDP\_per$).

$2.$ Industrial structure. Different industrial structures reflect the different endowments and comparative advantages of cities. From the perspective of urban evolution, digital economy development is also a process of industrial structure adjustment and optimization development. For the digital economy, the role of primary, secondary and tertiary industries is marginal increasing, so refer to the way of \citet{WangRenkoulaolinghuachanyejiegou2015} and others to construct the industrial structure upgrading index, whose formula is shown below :

\begin{equation}\label{2}		 
	Industrial\_structure=\sum_{1}^{3}{{Indust}_i\times i(1\le i\le3)}
\end{equation}

 ${Indust}_i$ denotes the share of the ith industry.

$ 3. $Internet penetration. The development of the digital economy relies on Internet technology. In this paper, we use the year-end population share of Internet users to measure the Internet penetration rate.

$ 4. $Government investment in research. The digital economy requires a technological dividend. Government intervention can influence urban planning and investment directions, and provide policy guarantees for new economic development. In this paper, we use the number of internal funding for research as government research investment ($RD\_capital$).

	\subsection{Data source}
	
The city data involved in this paper are mainly from the\textit{ Zhongguo tongji nianjian [China City Statistical Yearbook]},\textit{ Quyu tongji nianjian[China Regional Statistical Yearbook]}, with individual sample missing values combined with local statistical yearbooks, statistical bulletins or complemented by linear interpolation method. To maximize the construction of balanced panel data, this paper takes 282 prefecture-level cities, 136 smart city pilot cities and 70 national innovation pilot cities (except 4 municipalities directly under the central government and 2 county-level cities) from 2011 to 2020 as the objects of investigation. Considering the special characteristics of megacities, excluding four municipalities directly under the central government and Shenzhen.\footnote{The tabular results of the descriptive statistics are attached in the appendix (attached to the end of the paper after the references)}.

	\section{Empirical Analysis}	
	\subsection{ Basic regression }
	Table 2 reports the regression results of the policy effects of the dual pilot policy of smart cities and national innovation cities on the development of the digital economy. Where column (1) does not include control variables and uses random effects. Column (2) is the estimation result without adding control variables but controlling for city and year fixed effects, while columns (3) and (4) are the estimation results with control variables added to columns (1) and (2), respectively. The results show that the regression coefficients of ${Wtreat}_{i,t}$ are significantly positive regardless of the scenario, indicating to some extent that the dual pilot policy significantly enhances the level of digital economy development in cities.
\begin{table}[h]
		\caption{ Basic regression }
	\begin{center}
		\begin{tabular}{ccccc}
	\hline
	& (1)       & (2)       & (3)         & (4)       \\
	VARIABLES             & DEI       & DEI       & DEI         & DEI       \\ \hline
	\textit{Wtreatti}              & 9.2361*** & 2.3980*** & 4.0081***   & 2.2347*** \\
	& (34.0143) & (4.2745)  & (16.0445)   & (4.4574)  \\
	\textit{GDP\_per }             &           &           & 0.2043      & 0.6657*** \\
	&           &           & (1.0891)    & (2.9153)  \\
	\textit{industrial\_structure} &           &           & 0.0878***   & 0.0093    \\
	&           &           & (11.9444)   & (0.6964)  \\
	\textit{Internet }             &           &           & 5.0682***   & 4.9324*** \\
	&           &           & (9.1053)    & (5.4547)  \\
	\textit{RD\_capital}&           &           & 0.0000***   & 0.0000    \\
	&           &           & (25.5693)   & (0.7816)  \\
	\textit{Constant}    & 4.2689*** & 3.1438*** & -19.4775*** & -6.6347   \\
	& (41.2810) & (28.0375) & (-9.2895)   & (-1.6226) \\ \hline
  \textit{Observations } & 2,820     & 2,820     & 2,820       & 2,820     \\
	\textit{R-squared }  & 0.2911    & 0.4832    & 0.5705      & 0.5301    \\
	\textit{city.fe}      & no        & yes       & no          & yes       \\
	\textit{year.fe }     & no        & yes       & no          & yes       \\ \hline
\end{tabular}
	\end{center}

\end{table}

	\subsection{Parallel trend test and dynamic effects analysis}
	The premise of using the multi-temporal DID model is that the experimental group and the control group maintain the same trend of change before the policy occurs, i.e., the parallel trend test hypothesis is satisfied. Since the pilot cities receive policy shocks at different times, a time dummy variable cannot simply be set for a particular year as the threshold for policy occurrence, but a dummy variable for the relative time value of the implementation of the pilot policy in innovative cities needs to be set for each pilot city.

In this paper, we construct the following equation for parallel trend test as follows:
	
	\begin{equation}\label{3}
		\begin{split}		
{DEI}_{i,t}&=\alpha+\sum_{0}^{2}{\beta_i{Beforei}_{i,t}}+\beta_i{Current}_{i,t}+\sum_{0}^{6}{\beta_i{Afteri}_{i,t}}    \\	
			& +\gamma{Control_Var}_{i,t}+CityFE+YearFE+\varepsilon_{i,t}
		\end{split}
	\end{equation}

where the time dummy variables are the observations for the n years before, the current year, and the n years after each city is established as a pilot city. The dummy variables for non-pilot cities are all 0. Since the observation period of this paper is 2011-2020, while the year of policy implementation for the first pilot innovative cities is 2010 and the year of policy implementation for the first smart cities is 2013.

\begin{figure}[h]
\begin{center}
		\includegraphics[width=0.6\textwidth]{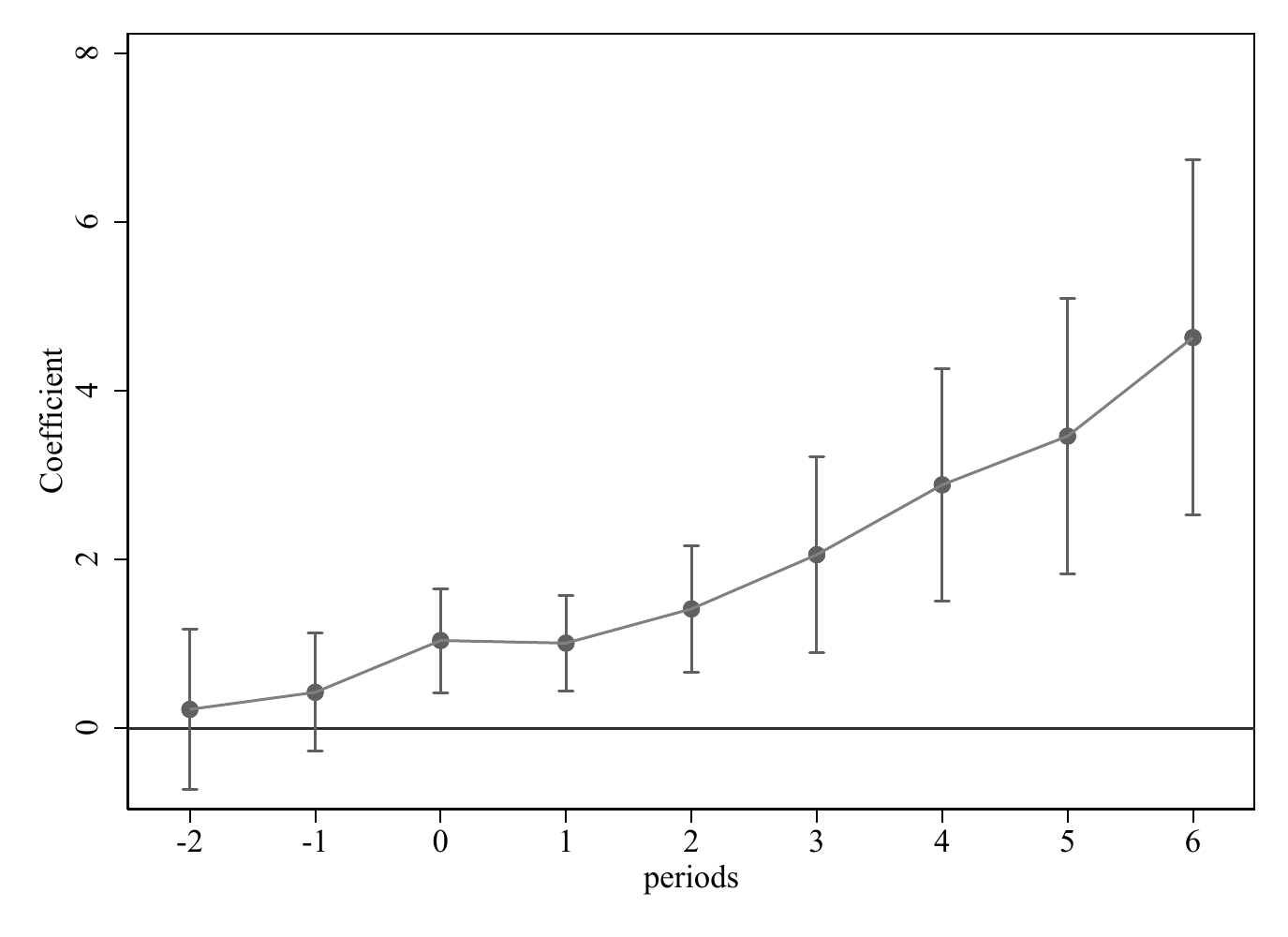}
	\caption{Parallel trend test} 
\end{center}
\end{figure}

The results show \footnote {more detailed graphs of the balance test are placed in the appendix at the end of the paper} that none of the coefficients of the relative time dummy variables before the onset of the policy are significant and have small values, which indicates that there is no significant difference in the level of digital economy development between the experimental group and the control group before the onset of the policy, i.e., the dual pilot policy is consistent with the parallel trend hypothesis.

Six years after the implementation of the pilot policy, the impact coefficient of the innovative city pilot policy is significantly positive and increasing, indicating that the innovative city pilot policy can produce policy effects that promote the improvement of the digital economy, but with a certain lag.

	\subsection{Robustness test}
	
	\subsubsection{Robustness test based on model setting}
	
	Multi-temporal propensity score matching-dual difference (PSM-DID) model. Although the double difference approach isolates the average treatment effect of the pilot policy, there may still be a selectivity bias problem in observing the research data because the national innovation city pilot policy is not a natural experiment in the strict sense. For this reason, this paper further conducts a robustness check based on a multi-temporal PSM-DID model. For the case that PSM is applied to cross-sectional data while DID is applied to panel data, there are two broad ideas in the literature at this stage to solve the problem: one is to construct a cross-sectional PSM, i.e., to treat panel data as cross-sectional data and then match them; the other is to match them period by period by referring to the approach adopted by 
 \citet{Bockerman2009Unemployment} 
 and others. In this paper, we use the panel data transformation method and the period-by-period matching method for propensity score matching in turn

The specific applications are:

$ 1. $ Set GDP per capita, industrial structure, Internet penetration rate, and government research support are set as matching variables. 

$ 2. $ Two datasets are obtained in two ways respectively: constructing a cross-sectional PSM, i.e., directly applying the nearest neighbor matching method to find the optimal control group that meets the common support condition for all innovative pilot cities and eliminating the non-common support part to obtain a new dataset; using the period-by-period matching method, which mainly matches the city samples year by year, and then merging the matched data of each year vertically into one dataset to generate panel data needed for regression.

 $ 3. $ Balance tests were conducted on the two matched data sets and the matching effects were analyzed. 
 
 $ 4. $ Re-estimate the impact effect of the city's dual pilot policy on the digital economy using a multi-temporal DID approach.

	Columns (1) and (2) in Table 3 report the regression results of the multi-temporal PSM-DID under the two methods. The results show that the coefficients of ${Wtreat}_{i,t}$ are still significantly positive and not substantially different from the baseline regression results, indicating to some extent that the effect of the dual pilot policy on enhancing the level of digital economy development is robust.

	\subsubsection{Replacing digital economy metrics}
 	Both the entropy value method and the principal component method are objectively assigned when calculating the indicator weights, and in this paper, we continue to use the original indicators as data, and the digital economy is analyzed by the principal component analysis method, and the newly generated digital economy is the variable DEI2. The regression results are shown in column (4) of Table 3. From the results, it can be seen that the coefficient of ${Wtreat}_{i,t}$ is significantly positive regardless of the metric used, and the results are robust.

 	\subsubsection{Exclude no policy time}
The national entrepreneurial city pilot policy introduced in 2010 and the smart city pilot policy established in three batches after 2012 are closely related to this paper during the examination period. Therefore, the data from 2011 to 2013 are removed from the baseline regression model in this paper. The results are shown in columns (5) and (6) in Table 3. From the estimation results, it can be found that the dummy variable coefficients are still significantly positive after controlling for the time horizon, and the results are still robust.
	
	\begin{table}[h]
		\caption{Robustness test}
\begin{center}
			\begin{tabular}{cccccc}
			\hline
			& (1)       & (2)       & (3)       & (4)       & (5)            \\
			VARIABLES    & DEI(Year-by-year)   & DEI(Cross-section)   & DEI-time  & DEI       & DEI2           \\ \hline
			\textit{Wtreatti }    & 1.7911*** & 1.8747*** & 1.2774*** & 2.2347*** & 50,294.2173*** \\
			& (18.0663) & (4.2917)  & (4.7068)  & (4.4574)  & (3.1256)       \\ \hline
		\textit{	Observations} & 1920      & 2493      & 1,692     & 2,820     & 2,820          \\
		\textit{	R-squared}    & 0.6944    & 0.5372    & 0.5883    & 0.5301    & 0.4781         \\
			\textit{city.fe  }     & yes       & yes       & yes       & yes       & yes            \\ 
			\textit{year.fe }      & yes       & yes       & yes       & yes       & yes  \\ \hline       
		\end{tabular}
\end{center}
	\end{table}

	\subsubsection{Placebo test}
	
Although a large number of city characteristic variables have been controlled for in this paper in a quasi-natural experiment, there may still be some unobserved city characteristic factors that make the evaluation results of the innovative city pilot policy influenced. If the model is a simultaneous point DID, referring to ~\citet {Liu2015Firm}, a random selection of cities equal to the number of real pilot cities from all sample cities is sufficient as the treatment group .

\begin{figure}[h]
	\begin{center}
		\includegraphics[width=0.8\textwidth]{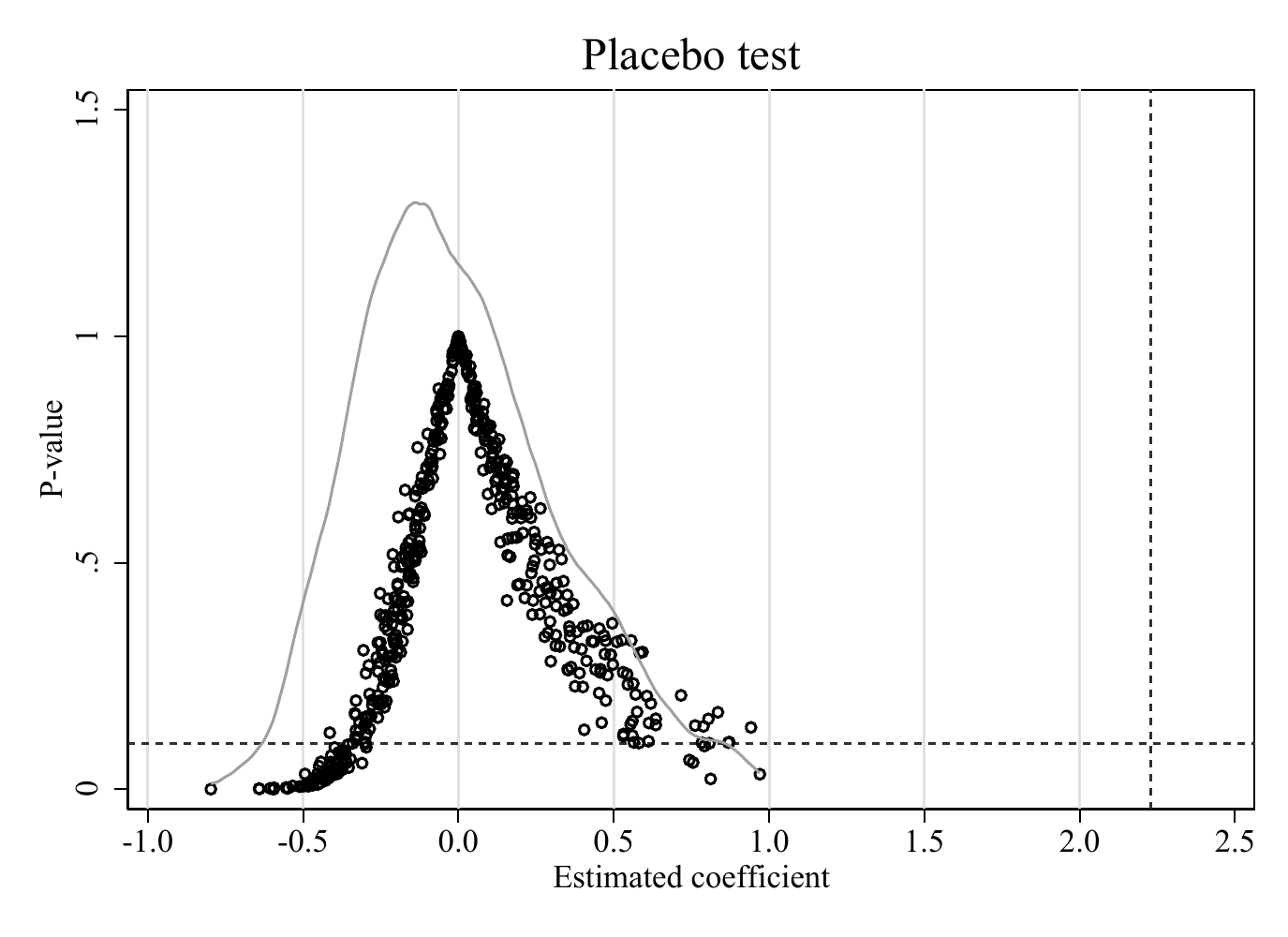}
		\caption{  Placebo test}
	\end{center} 
\end{figure}

 However, due to the differences in the policy shock times of pilot cities in the multi-temporal DID, both the pseudo-treatment group dummy variable $Grouprandom$ and the pseudo-policy shock dummy variable $Postrandom$ need to be randomly generated, i.e., a sample period is randomly selected for each sample object as its policy time. Based on this, to further ensure the robustness of the estimation results, the following placebo test is used in this paper:To ensure that the policy cannot have an actual impact on the development of the digital economy, Stata software is used to construct 500 random shocks of the pseudo-innovative city pilot policy on 282 sample cities, with 72 cities randomly selected as the experimental group each time, and the policy time is given randomly to obtain the 500-group dummy variables $Policyrandom$ (i.e., $Grouprandom\times Postrandom$); the kernel density of 500 $\beta random$ and its p-value distribution are presented in the figure.
	
The results showed that the $\beta random$ generated during the randomization process were mainly concentrated around 0 and the p-values were mostly higher than 0.1, while the estimated coefficient of the actual policy was 0.2347, which was significantly different from the placebo test results. This also suggests, to some extent, that the quantitative assessment results in this paper were not significantly influenced by this potential factor. The results are robust.
	
	\subsection{Heterogeneity analysis}
	
The previous analysis shows that the construction of national innovative cities and smart cities can significantly accelerate the development of urban digital economy, but because of the uneven development of various regions in China, and the location condition is a key factor affecting the construction of cities, cities with good location conditions have the advantages of innate resource endowment, easier access to factors, and lower transaction costs, so the level of city construction varies from region to region. This section groups city data based on four major economic regions and Hu  line to test regional heterogeneity.

\begin{table}[h]
	\caption{Heterogeneity test group regression}
\begin{center}
		\begin{tabular}{c|cccc|cc}
		\hline
		Variable Name & \multicolumn{4}{c|}{Four Economic Regions}                  & \multicolumn{2}{c}{Hu Line} \\ \hline
		Sequence& (1)       & (2)       & (3)      & (4)       & (5)         & (6)        \\
		VARIABLES    & EAST      & MIDDLE    & WEST     & NOR\_EAST & HU\_LINE1   & HU\_LINE2  \\ \hline
	\textit{	Wtreatti }    & 1.7502*** & 1.0369**  & 0.1844   & 0.8822    & 1.0425***   & 0.5194*    \\
		& (3.2263)  & (2.5303)  & (0.2589) & (1.2012)  & (2.9901)    & (1.7525)   \\
		Constant     & -0.5093   & -12.0601  & 0.5634   & -5.4052   & -4.7418     & -3.9322    \\
		& (-0.1759) & (-1.0660) & (0.0756) & (-1.5480) & (-1.2752)   & (-0.6410)  \\ \hline
	\textit{	Observations} & 340       & 820       & 800      & 860       & 2,570       & 250        \\
		\textit{R-squared}    & 0.8185    & 0.7827    & 0.7297   & 0.6679    & 0.7043      & 0.8175     \\
		\textit{Number of id}   & 34        & 82        & 80       & 86        & 257         & 25         \\ \hline
	\end{tabular}
\end{center}
\end{table}

According to Table 4, there is regional heterogeneity in the effect of the dual pilot city policy on the development of digital economy. As seen from model (1)(2)(3)(4):the construction of smart cities and innovation cities in the eastern and central regions significantly accelerated the level of digital economy development, but the effect in the eastern and northeastern regions was not significant. The main reasons for this are:First, the relatively serious aging of the population in the eastern and western regions leads to a lack of talent gathering effect, which ultimately hinders industrial upgrading ~\citep{ZhuoRenkoulaolinghaquyuchuangx2018}. Second, the large population outflow from the northeast reduces the accumulation of human capital in the region, and the low level of human capital will limit the market of the digital economy in the region. Meanwhile, the digital economy relies on logistics networks and the Internet, and industrial development in the northeast and western regions is trapped by the construction of hardware infrastructure. This is also related to topography; for example, Yunnan's highways cost more than other provinces because they are built through more mountains and cost more.

The pilot policy still has the policy effect of enhancing entrepreneurial activity for cities on the northwest side of the Hu Line\footnote{It stretches from the city of Heihe in northeast to Tengchong in south, diagonally across China.}, but this policy effect is weaker than that on the Hu Line  and its southeast side. The agglomeration phenomenon generated by the spatial flow of factor resources is essentially the result of the combined effect of urban attractiveness and exclusion. Although the pilot city policy has tried to enhance the attractiveness of the pilot cities to innovation factors, due to the level of economic development, entrepreneurial infrastructure, and market environment, the flow of resources such as population has long shown a preference for "southeastward flight" ~\citet{BaiChuangxingqudongzhengce2022}. Compared with cities east of the Hu Huanyong line, cities to the west have relatively weaker ability to gather innovation factors and are at a relative disadvantage in attracting factor inflows and optimizing resource allocation, thus affecting to a certain extent the full utilization of pilot policies in cities in this region.

\subsection{Mechanism Testing}

Based on the previous theoretical analysis, in order to test the mechanism, the following mediating effect model is constructed in this paper:
\begin{equation}\label{4}		 {Inter\_var}_{i,t}=\alpha+\varphi{Wtreat}_{i,t}+\gamma{Control_Var}_{i,t}+CityFE+YearFE+\varepsilon_{i,t}
\end{equation}
\begin{equation}\label{5}		 {DEI}_{i,t}=\alpha+\theta{Wtreat}_{i,t}+\delta{Inter\_var}_{i,t}+\gamma{Control_Var}_{i,t}+CityFE+YearFE+\varepsilon_{i,t}
\end{equation}

Among them, ${Inter\_var}_{i,t}$ of equation (4) is the mediating variable, and three variables reflecting the investment agglomeration effect (VCPE), talent agglomeration effect (Talents), and technology agglomeration effect (Technology) are used in turn for replacement. Other variables are consistent with the previous section. If the coefficients $ \varphi$ and $ \delta$ are both significant, the mediating effect holds. Further, if $ \theta $ is also significant and has the same sign as $\varphi\times\delta$, it means that ${Inter\_var}_{i,t}$ has a partial mediating effect and its contribution to the total effect is $\varphi\times\delta/(\varphi\times\delta+\theta )$.

\textbf{1. investment clustering effect (VCPE).} The indicator used is the degree of financial support, which is estimated by the ratio of the amount of financial institutions to local gdp at the end of the year. The regression results are shown in columns (1) and (2) in Table 4. The positive coefficient of the variable ${Wtreat}_{i,t}$ in column (1) indicates that the implementation of the national innovative city pilot policy significantly increased the investment level in the pilot cities. The pilot policy of innovative cities not only releases favorable policy information to investors, but also the pilot cities introduce relevant policies to support venture capital, thus enhancing investors' willingness to invest in venture capital for entrepreneurs in the pilot cities and positively influencing the improvement of venture capital level in the pilot cities, and the contribution of this effect to the total effect can be found to be about 5.67\% by combining the estimated coefficients.

\textbf{2. Talent Aggregation Effect (Talents).} The development of digital economy cannot be effectively supported by high knowledge talents. Given the availability of city-level data, this paper uses the sum of the number of employees in the scientific research, technical services and geological exploration industry and the number of employees in the information transmission, computer services and software industry, the number of employees in the financial industry, and the number of employees in the education industry as a proportion of the total number of employees in the city to characterize the agglomeration level of talents in the city in an approximate manner. Since the knowledge and technology level of employees in these industries is relatively high compared to other industries, it is reasonable and feasible to use this indicator to measure the talent level of a city when the data of common indicators such as the average years of education at the city level are difficult to obtain. From the regression results in column (3) of Table 4, the coefficient of ${Wtreat}_{i,t}$ is significantly positive at the level of 1\%, indicating that the pilot policy of innovative cities has exerted a talent clustering effect, and the results in column (4) of Table 4 show that the coefficients of ${Wtreat}_{i,t}$ and Talents are both significantly positive, indicating that talent clustering has to a certain extent enriched the potential entrepreneurial subjects of entrepreneurial activities and promoted The contribution of this mediating effect to the total effect is calculated to be about 40.9\%. Accordingly, the mechanism of the pilot innovative city policy acting on entrepreneurial activity through talent clustering effect is verified.

\textbf{3. Technology agglomeration effect(Technology).} In this paper, we use "(government investment in science and technology/total government budget expenditure) × 0.5 + standardized value of patent grants per capita × 0.5" to construct the technology agglomeration index \footnote{using z-score standardization, i.e. standard deviation standardization for patent applications per capita.} . The regression results are shown in columns (5) and (6) of Table 4. Among them, the regression coefficient of ${Wtreat}_{i,t}$ in column (1) is positive, which indicates that the pilot policy has effectively improved the output of technological innovation results in the city and contributed to the technological agglomeration in the city. Combined with the results in column (2), it can be seen that the technology agglomeration effect generated by the pilot policy further drives up the level of digital economy development, and the contribution of this part of the mediating effect to the total effect is about 45.57\%.

\begin{table}[H]
	\caption{ Mechanism Testing}
	\begin{tabular}{c|cc|cc|cc}
		\hline
Sequence          & (1)      & (2)       & (3)        & (4)       & (5)        & (6)       \\ \hline
VARIABLES    & vcpe     & DEI       & talents    & DEI       & technology & DEI       \\ \hline
\textit{\textit{Wtreatti }}    & 0.0034*  & 0.9061*** & 35.8273*** & 0.5674**  & 0.0106*    & 0.9394*** \\
& (1.7224) & (2.9913)  & (2.6566)   & (2.5487)  & (2.1949)   & (3.0398)  \\
\textit{Finance\_g}   &          & 16.0090*  &            &           &            &           \\
&          & (1.7142)  &            &           &            &           \\
\textit{Talents }     &          &           &            & 0.0110*** &            &           \\
&          &           &            & (5.5610)  &            &           \\
\textit{technology}   &          &           &            &           &            & 1.9172**  \\
&          &           &            &           &            & (2.4554)  \\ \hline
\textit{Observations} & 2,820    & 2,820     & 2,820      & 2,820     & 2,820      & 2,820     \\
\textit{R-squared}    & 0.3749   & 0.7140    & 0.1412     & 0.7558    & 0.1294     & 0.7405    \\ \hline
	\end{tabular}
\end{table}

\begin{center}
	\section{Conclusions and Policy Implications}
\end{center}

The digital economy is leading China's quality change, efficiency change, and power change, and has become a new driving force to promote the development of common prosperity ~\citep{kolesnikov2020global}. Smart cities can also promote common prosperity through the path of urban construction. In the context of smart city and national innovation city construction, its panel data of 282 prefecture-level and above cities in China from 2011 to 2020, adopting a multi-period DID approach (Time-varyingDID), empirically analyzed the impact effect and mechanism of the development of digital economy on common wealth. The results find that the dual pilot policy facilitates digital economy development, and this finding still holds after robustness tests such as placebo tests and controlling for shocks from other similar policies. Further research shows that the impact effect of digital economy development on shared prosperity is more pronounced in cities with higher population density and advantageous geographical location. Based on the above findings, this paper proposes the following policy implications.

1. Pay attention to the important role of city policies in boosting cities to gather venture capital, talent, technology and policies, and deeply unblock the transmission channels of this policy to promote urban economic development. Specifically, in order to give full play to the agglomeration function of the policy to help the digital economy, its optimization can focus on the following aspects: strengthen the financial investment attraction of the relevant institutional construction, the establishment of a sound institutional mechanism and platform construction conducive to investment entry and exit; strengthen the guidance of digital economy investment; further play the advantages of policy agglomeration, multi-pronged approach to optimize the city's business environment and reduce transaction costs.

2. Through the refinement and re-planning of the national urban policy, the policy is enhanced to be targeted to different cities or industries to avoid the "one-size-fits-all" policy. The heterogeneity test results of this paper show that there is still a "Matthew effect" among cities with different geographical locations. Therefore, in the process of building innovative cities in the future, we should strengthen the coordination among different cities and regions, formulate and implement innovation-driven policies according to local conditions, and promote the coordinated development of urban digital economy.

3. Strengthen infrastructure construction and policy coordination. In the heterogeneity analysis, one of the reasons for the huge difference in the level of digital economy development is the difference in infrastructure, with the western region and the northeastern region relatively lagging behind in the construction of logistics network and Internet facilities. High-speed broadband network is one of the infrastructures for the development of digital economy. By building a high-speed broadband network, it can increase the speed of network connection and help improve the network coverage. This will provide better information infrastructure for the development of digital economy. Attention should be paid to the construction of logistics networks in the northeast and western regions increasing investment in hardware in relatively backward regions. At the same time there is a coordination effect between policies, and urban dual policy pilots should be actively promoted. Bridge the hardware gap through smart cities, and then enhance innovation capacity through national innovation city policy.

\bibliographystyle{apacite}
\bibliography{ref}

\begin{thebibliography}{}

\bibitem [\protect \citeauthoryear {%
Angelidou%
}{%
Angelidou%
}{%
{\protect \APACyear {2014}}%
}]{%
Angelidou2014Smart}
\APACinsertmetastar {%
Angelidou2014Smart}%
\begin{APACrefauthors}%
Angelidou, M.%
\end{APACrefauthors}%
\unskip\
\newblock
\APACrefYearMonthDay{2014}{7}{}.
\newblock
{\BBOQ}\APACrefatitle {Smart city policies: A spatial approach} {Smart city
  policies: A spatial approach}.{\BBCQ}
\newblock
\APACjournalVolNumPages{Cities}{41}{}{S3--S11}.
\newblock
\begin{APACrefDOI} \doi{10.1016/j.cities.2014.06.007} \end{APACrefDOI}
\PrintBackRefs{\CurrentBib}

\bibitem [\protect \citeauthoryear {%
Bai%
, Zhang%
\BCBL {}\ \BBA {} Bian%
}{%
Bai%
\ \protect \BOthers {.}}{%
{\protect \APACyear {2022}}%
}]{%
BaiChuangxingqudongzhengce2022}
\APACinsertmetastar {%
BaiChuangxingqudongzhengce2022}%
\begin{APACrefauthors}%
Bai, J.%
, Zhang, Y.%
\BCBL {}\ \BBA {} Bian, Y.%
\end{APACrefauthors}%
\unskip\
\newblock
\APACrefYearMonthDay{2022}{}{}.
\newblock
{\BBOQ}\APACrefatitle {Chuangxing qudong zhengce shifou tisheng chengshi
  chuangye huoyuedu\textemdash\textemdash laizi guojia chuangxingxing chengshi
  shidian zhengce de jingyan zhengju[Does Innovation-driven Policy Increase
  Entrepreneurial Activity in Cities \textemdash\textemdash\textemdash Evidence
  from the National Innovative City Pilot Policy]} {Chuangxing qudong zhengce
  shifou tisheng chengshi chuangye huoyuedu\textemdash\textemdash laizi guojia
  chuangxingxing chengshi shidian zhengce de jingyan zhengju[does
  innovation-driven policy increase entrepreneurial activity in cities
  \textemdash\textemdash\textemdash evidence from the national innovative city
  pilot policy]}.{\BBCQ}
\newblock
\APACjournalVolNumPages{Zhongguo gongye jingji}{}{6}{61--78}.
\newblock
\begin{APACrefDOI} \doi{10.19581/j.cnki.ciejournal.2022.06.016}
  \end{APACrefDOI}
\PrintBackRefs{\CurrentBib}

\bibitem [\protect \citeauthoryear {%
Bockerman%
\ \BBA {} Ilmakunnas%
}{%
Bockerman%
\ \BBA {} Ilmakunnas%
}{%
{\protect \APACyear {2009}}%
}]{%
Bockerman2009Unemployment}
\APACinsertmetastar {%
Bockerman2009Unemployment}%
\begin{APACrefauthors}%
Bockerman, P.%
\BCBT {}\ \BBA {} Ilmakunnas, P.%
\end{APACrefauthors}%
\unskip\
\newblock
\APACrefYearMonthDay{2009}{2}{}.
\newblock
{\BBOQ}\APACrefatitle {Unemployment and self-assessed health: evidence from
  panel data} {Unemployment and self-assessed health: evidence from panel
  data}.{\BBCQ}
\newblock
\APACjournalVolNumPages{Health Economics}{18}{2}{161--179}.
\newblock
\begin{APACrefDOI} \doi{10.1002/hec.1361} \end{APACrefDOI}
\PrintBackRefs{\CurrentBib}

\bibitem [\protect \citeauthoryear {%
Caragliu%
\ \BBA {} Del~Bo%
}{%
Caragliu%
\ \BBA {} Del~Bo%
}{%
{\protect \APACyear {2019}}%
}]{%
Caragliu2019Smart}
\APACinsertmetastar {%
Caragliu2019Smart}%
\begin{APACrefauthors}%
Caragliu, A.%
\BCBT {}\ \BBA {} Del~Bo, C\BPBI F.%
\end{APACrefauthors}%
\unskip\
\newblock
\APACrefYearMonthDay{2019}{5}{}.
\newblock
{\BBOQ}\APACrefatitle {Smart innovative cities: The impact of Smart City
  policies on urban innovation} {Smart innovative cities: The impact of smart
  city policies on urban innovation}.{\BBCQ}
\newblock
\APACjournalVolNumPages{Technological Forecasting and Social
  Change}{142}{}{373--383}.
\newblock
\begin{APACrefDOI} \doi{10.1016/j.techfore.2018.07.022} \end{APACrefDOI}
\PrintBackRefs{\CurrentBib}

\bibitem [\protect \citeauthoryear {%
Carlsson%
}{%
Carlsson%
}{%
{\protect \APACyear {2004}}%
}]{%
Carlsson2004Digital}
\APACinsertmetastar {%
Carlsson2004Digital}%
\begin{APACrefauthors}%
Carlsson, B.%
\end{APACrefauthors}%
\unskip\
\newblock
\APACrefYearMonthDay{2004}{9}{}.
\newblock
{\BBOQ}\APACrefatitle {The Digital Economy: what is new and what is not?} {The
  digital economy: what is new and what is not?}{\BBCQ}
\newblock
\APACjournalVolNumPages{Structural Change and Economic
  Dynamics}{15}{3}{245--264}.
\newblock
\begin{APACrefDOI} \doi{10.1016/j.strueco.2004.02.001} \end{APACrefDOI}
\PrintBackRefs{\CurrentBib}

\bibitem [\protect \citeauthoryear {%
Gao%
\ \BBA {} Yuan%
}{%
Gao%
\ \BBA {} Yuan%
}{%
{\protect \APACyear {2022}}%
}]{%
Gao2022Government}
\APACinsertmetastar {%
Gao2022Government}%
\begin{APACrefauthors}%
Gao, K.%
\BCBT {}\ \BBA {} Yuan, Y.%
\end{APACrefauthors}%
\unskip\
\newblock
\APACrefYearMonthDay{2022}{8}{}.
\newblock
{\BBOQ}\APACrefatitle {Government intervention, spillover effect and urban
  innovation performance: Empirical evidence from national innovative city
  pilot policy in China} {Government intervention, spillover effect and urban
  innovation performance: Empirical evidence from national innovative city
  pilot policy in china}.{\BBCQ}
\newblock
\APACjournalVolNumPages{Technology in Society}{70}{}{102035}.
\newblock
\begin{APACrefDOI} \doi{10.1016/j.techsoc.2022.102035} \end{APACrefDOI}
\PrintBackRefs{\CurrentBib}

\bibitem [\protect \citeauthoryear {%
Guo%
\ \protect \BOthers {.}}{%
Guo%
\ \protect \BOthers {.}}{%
{\protect \APACyear {2020}}%
}]{%
Guo2020Cedu}
\APACinsertmetastar {%
Guo2020Cedu}%
\begin{APACrefauthors}%
Guo, F.%
, Wang, J.%
, Wang, F.%
, Kong, T.%
, Zhang, X.%
\BCBL {}\ \BBA {} Cheng, Z.%
\end{APACrefauthors}%
\unskip\
\newblock
\APACrefYearMonthDay{2020}{}{}.
\newblock
{\BBOQ}\APACrefatitle {Cedu Zhongguo shuzi puhui jingrong fazhan:zhishu bianzhi
  yu kongjian tezheng [Measuring China's Digital Financial Inclusion: Index
  Compilation and Spatial Characteristics]} {Cedu zhongguo shuzi puhui jingrong
  fazhan:zhishu bianzhi yu kongjian tezheng [measuring china's digital
  financial inclusion: Index compilation and spatial characteristics]}.{\BBCQ}
\newblock
\APACjournalVolNumPages{Jingji xue(jikan)}{19}{4}{1401--1418}.
\newblock
\begin{APACrefDOI} \doi{10.13821/j.cnki.ceq.2020.03.12} \end{APACrefDOI}
\PrintBackRefs{\CurrentBib}

\bibitem [\protect \citeauthoryear {%
Jiang%
}{%
Jiang%
}{%
{\protect \APACyear {2020}}%
}]{%
Jiang2020Digital}
\APACinsertmetastar {%
Jiang2020Digital}%
\begin{APACrefauthors}%
Jiang, X.%
\end{APACrefauthors}%
\unskip\
\newblock
\APACrefYearMonthDay{2020}{10}{1}.
\newblock
{\BBOQ}\APACrefatitle {Digital economy in the post-pandemic era} {Digital
  economy in the post-pandemic era}.{\BBCQ}
\newblock
\APACjournalVolNumPages{Journal of Chinese Economic and Business
  Studies}{18}{4}{333--339}.
\newblock
\begin{APACrefDOI} \doi{10.1080/14765284.2020.1855066} \end{APACrefDOI}
\PrintBackRefs{\CurrentBib}

\bibitem [\protect \citeauthoryear {%
Kolesnikov%
, Zernova%
, Degtyareva%
, Panko%
\BCBL {}\ \BBA {} Sigidov%
}{%
Kolesnikov%
\ \protect \BOthers {.}}{%
{\protect \APACyear {2020}}%
}]{%
kolesnikov2020global}
\APACinsertmetastar {%
kolesnikov2020global}%
\begin{APACrefauthors}%
Kolesnikov, A.%
, Zernova, L.%
, Degtyareva, V.%
, Panko, I\BPBI V.%
\BCBL {}\ \BBA {} Sigidov, Y\BPBI I.%
\end{APACrefauthors}%
\unskip\
\newblock
\APACrefYearMonthDay{2020}{}{}.
\newblock
{\BBOQ}\APACrefatitle {Global trends of the digital economy development}
  {Global trends of the digital economy development}.{\BBCQ}
\newblock
\APACjournalVolNumPages{Opci{\'o}n: Revista de Ciencias Humanas y
  Sociales}{}{26}{523--540}.
\PrintBackRefs{\CurrentBib}

\bibitem [\protect \citeauthoryear {%
Li%
, Li%
, Ma%
, Zheng%
\BCBL {}\ \BBA {} Pan%
}{%
Li%
\ \protect \BOthers {.}}{%
{\protect \APACyear {2022}}%
}]{%
Li2022Does}
\APACinsertmetastar {%
Li2022Does}%
\begin{APACrefauthors}%
Li, L.%
, Li, M.%
, Ma, S.%
, Zheng, Y.%
\BCBL {}\ \BBA {} Pan, C.%
\end{APACrefauthors}%
\unskip\
\newblock
\APACrefYearMonthDay{2022}{9}{}.
\newblock
{\BBOQ}\APACrefatitle {Does the construction of innovative cities promote urban
  green innovation?} {Does the construction of innovative cities promote urban
  green innovation?}{\BBCQ}
\newblock
\APACjournalVolNumPages{Journal of Environmental Management}{318}{}{115605}.
\newblock
\begin{APACrefDOI} \doi{10.1016/j.jenvman.2022.115605} \end{APACrefDOI}
\PrintBackRefs{\CurrentBib}

\bibitem [\protect \citeauthoryear {%
Liu%
\ \BBA {} Lu%
}{%
Liu%
\ \BBA {} Lu%
}{%
{\protect \APACyear {2015}}%
}]{%
Liu2015Firm}
\APACinsertmetastar {%
Liu2015Firm}%
\begin{APACrefauthors}%
Liu, Q.%
\BCBT {}\ \BBA {} Lu, Y.%
\end{APACrefauthors}%
\unskip\
\newblock
\APACrefYearMonthDay{2015}{11}{}.
\newblock
{\BBOQ}\APACrefatitle {Firm investment and exporting: Evidence from China's
  value-added tax reform} {Firm investment and exporting: Evidence from china's
  value-added tax reform}.{\BBCQ}
\newblock
\APACjournalVolNumPages{Journal of International Economics}{97}{2}{392--403}.
\newblock
\begin{APACrefDOI} \doi{10.1016/j.jinteco.2015.07.003} \end{APACrefDOI}
\PrintBackRefs{\CurrentBib}

\bibitem [\protect \citeauthoryear {%
Mesenbourg%
}{%
Mesenbourg%
}{%
{\protect \APACyear {2001}}%
}]{%
mesenbourg2001measuring}
\APACinsertmetastar {%
mesenbourg2001measuring}%
\begin{APACrefauthors}%
Mesenbourg, T\BPBI L.%
\end{APACrefauthors}%
\unskip\
\newblock
\APACrefYearMonthDay{2001}{}{}.
\newblock
{\BBOQ}\APACrefatitle {Measuring the digital economy} {Measuring the digital
  economy}.{\BBCQ}
\newblock
\APACjournalVolNumPages{US Bureau of the Census}{1}{}{1--19}.
\PrintBackRefs{\CurrentBib}

\bibitem [\protect \citeauthoryear {%
Strassner%
\ \BBA {} Nicholson%
}{%
Strassner%
\ \BBA {} Nicholson%
}{%
{\protect \APACyear {2020}}%
}]{%
Strassner2020Measuring}
\APACinsertmetastar {%
Strassner2020Measuring}%
\begin{APACrefauthors}%
Strassner, E\BPBI H.%
\BCBT {}\ \BBA {} Nicholson, J\BPBI R.%
\end{APACrefauthors}%
\unskip\
\newblock
\APACrefYearMonthDay{2020}{8}{26}.
\newblock
{\BBOQ}\APACrefatitle {Measuring the digital economy in the United States}
  {Measuring the digital economy in the united states}.{\BBCQ}
\newblock
\APACjournalVolNumPages{Statistical Journal of the IAOS}{36}{3}{647--655}.
\newblock
\begin{APACrefDOI} \doi{10.3233/SJI-200666} \end{APACrefDOI}
\PrintBackRefs{\CurrentBib}

\bibitem [\protect \citeauthoryear {%
S.~Wang%
, Wang%
, Wei%
, Wang%
\BCBL {}\ \BBA {} Fan%
}{%
S.~Wang%
\ \protect \BOthers {.}}{%
{\protect \APACyear {2021}}%
}]{%
Wang2021Collaborative}
\APACinsertmetastar {%
Wang2021Collaborative}%
\begin{APACrefauthors}%
Wang, S.%
, Wang, J.%
, Wei, C.%
, Wang, X.%
\BCBL {}\ \BBA {} Fan, F.%
\end{APACrefauthors}%
\unskip\
\newblock
\APACrefYearMonthDay{2021}{9}{}.
\newblock
{\BBOQ}\APACrefatitle {Collaborative innovation efficiency: From within cities
  to between cities—Empirical analysis based on innovative cities in China}
  {Collaborative innovation efficiency: From within cities to between
  cities—empirical analysis based on innovative cities in china}.{\BBCQ}
\newblock
\APACjournalVolNumPages{Growth and Change}{52}{3}{1330--1360}.
\newblock
\begin{APACrefDOI} \doi{10.1111/grow.12504} \end{APACrefDOI}
\PrintBackRefs{\CurrentBib}

\bibitem [\protect \citeauthoryear {%
W.~Wang%
, Liu%
\BCBL {}\ \BBA {} Peng%
}{%
W.~Wang%
\ \protect \BOthers {.}}{%
{\protect \APACyear {2015}}%
}]{%
WangRenkoulaolinghuachanyejiegou2015}
\APACinsertmetastar {%
WangRenkoulaolinghuachanyejiegou2015}%
\begin{APACrefauthors}%
Wang, W.%
, Liu, Y.%
\BCBL {}\ \BBA {} Peng, D.%
\end{APACrefauthors}%
\unskip\
\newblock
\APACrefYearMonthDay{2015}{}{}.
\newblock
{\BBOQ}\APACrefatitle {Renkou laolinghua de chanyejiegou shengjixiaoying
  yanjiu[Research on Effects of Population Aging on Industrial Upgrading]}
  {Renkou laolinghua de chanyejiegou shengjixiaoying yanjiu[research on effects
  of population aging on industrial upgrading]}.{\BBCQ}
\newblock
\APACjournalVolNumPages{Zhongguo gongye jingji}{}{11}{47--61}.
\newblock
\begin{APACrefDOI} \doi{10.19581/j.cnki.ciejournal.2015.11.004}
  \end{APACrefDOI}
\PrintBackRefs{\CurrentBib}

\bibitem [\protect \citeauthoryear {%
Zhao%
, Zhang%
\BCBL {}\ \BBA {} Liang%
}{%
Zhao%
\ \protect \BOthers {.}}{%
{\protect \APACyear {2020}}%
}]{%
Zhao2020Shuzi}
\APACinsertmetastar {%
Zhao2020Shuzi}%
\begin{APACrefauthors}%
Zhao, T.%
, Zhang, Z.%
\BCBL {}\ \BBA {} Liang, S.%
\end{APACrefauthors}%
\unskip\
\newblock
\APACrefYearMonthDay{2020}{}{}.
\newblock
{\BBOQ}\APACrefatitle {Shuzi jingji,chuangye huoyuedu yu gaozhiliang
  fazhan:laizi Zhongguo chengshi de jingyan zhengju [Digital Economy,
  Entrepreneurship, and High-Quality Economic Development:Empirical Evidence
  from Urban China]} {Shuzi jingji,chuangye huoyuedu yu gaozhiliang
  fazhan:laizi zhongguo chengshi de jingyan zhengju [digital economy,
  entrepreneurship, and high-quality economic development:empirical evidence
  from urban china]}.{\BBCQ}
\newblock
\APACjournalVolNumPages{Guanli shijie}{36}{10}{65--76}.
\newblock
\begin{APACrefDOI} \doi{10.19744/j.cnki.11-1235/f.2020.0154} \end{APACrefDOI}
\PrintBackRefs{\CurrentBib}

\bibitem [\protect \citeauthoryear {%
Zhuo%
\ \BBA {} Deng%
}{%
Zhuo%
\ \BBA {} Deng%
}{%
{\protect \APACyear {2018}}%
}]{%
ZhuoRenkoulaolinghaquyuchuangx2018}
\APACinsertmetastar {%
ZhuoRenkoulaolinghaquyuchuangx2018}%
\begin{APACrefauthors}%
Zhuo, C.%
\BCBT {}\ \BBA {} Deng, F.%
\end{APACrefauthors}%
\unskip\
\newblock
\APACrefYearMonthDay{2018}{}{}.
\newblock
{\BBOQ}\APACrefatitle {Renkou laolingha,quyuchuangx yu chanyejiegou
  shengji[Population Aging,Regional Innovation and the Upgrade of Industrial
  Structure]} {Renkou laolingha,quyuchuangx yu chanyejiegou shengji[population
  aging,regional innovation and the upgrade of industrial structure]}.{\BBCQ}
\newblock
\APACjournalVolNumPages{Renkou yu jingji}{}{1}{48--60}.
\PrintBackRefs{\CurrentBib}

\end{thebibliography}

\appendix

	\section{ Appendix}
\subsection*{Appendix I: Data situation}

	\begin{table}[h]
	\caption{ Data situation}
\begin{center}
	\begin{tabular}{cccccc}
		\hline
		Variable          & Obs  & Mean     & Std.Dev. & Min    & Max      \\ \hline
		\textit{DEI }              & 2820 & 5.6084   & 6.0295   & 0.1953  & 58.2121  \\
		\textit{Wtreatti}          & 2820 & 0.145     & .3522    & 0      & 1        \\
		\textit{GDP\_per }         & 2820 & 10.7005  & .5635    & 8.7729 & 12.5793  \\
	\textit{	industrial$\sim$e} & 2820 & 229.2122 & 13.8974  & 183.12 & 276.03   \\
		\textit{Internet}          & 2820 & 0.2348    & .1837    & 0.0035  & 1.4893   \\
	\textit{	RD\_capital }      & 2820 & 452000   & 1010000  & 3      & 1.88e+07 \\ \hline
	\end{tabular}
\end{center}
\end{table}

\subsection*{Appendix II: Principal component analysis for weights}

\begin{table}[h]
	\caption{Principal component method indicator weights}
		\begin{center}
			\begin{tabular}{cccccc}
		\hline
		 Indicator 1 & Indicator 2 &  Indicator 3 & Indicator 4 & Indicator 5 & Indicator 6 \\ \hline
		0.101   & 0.117     & 0.161  & 0.301    & 0.292   & 0.342   \\ \hline
	\end{tabular}
		\end{center}
\end{table}

The six indicators correspond to phone popularity,Internet users, Telecom Business,Digital Inclusive Finance,related practitioners and patents granted.
\subsection*{Appendix III: Balance test}
The following figure shows the results of the equilibrium test after matching the panel data as if they were cross-sectional data:

\begin{center}
	\begin{figure}[h]
		\begin{minipage}{.5\textwidth}
			\includegraphics[width=1\linewidth]{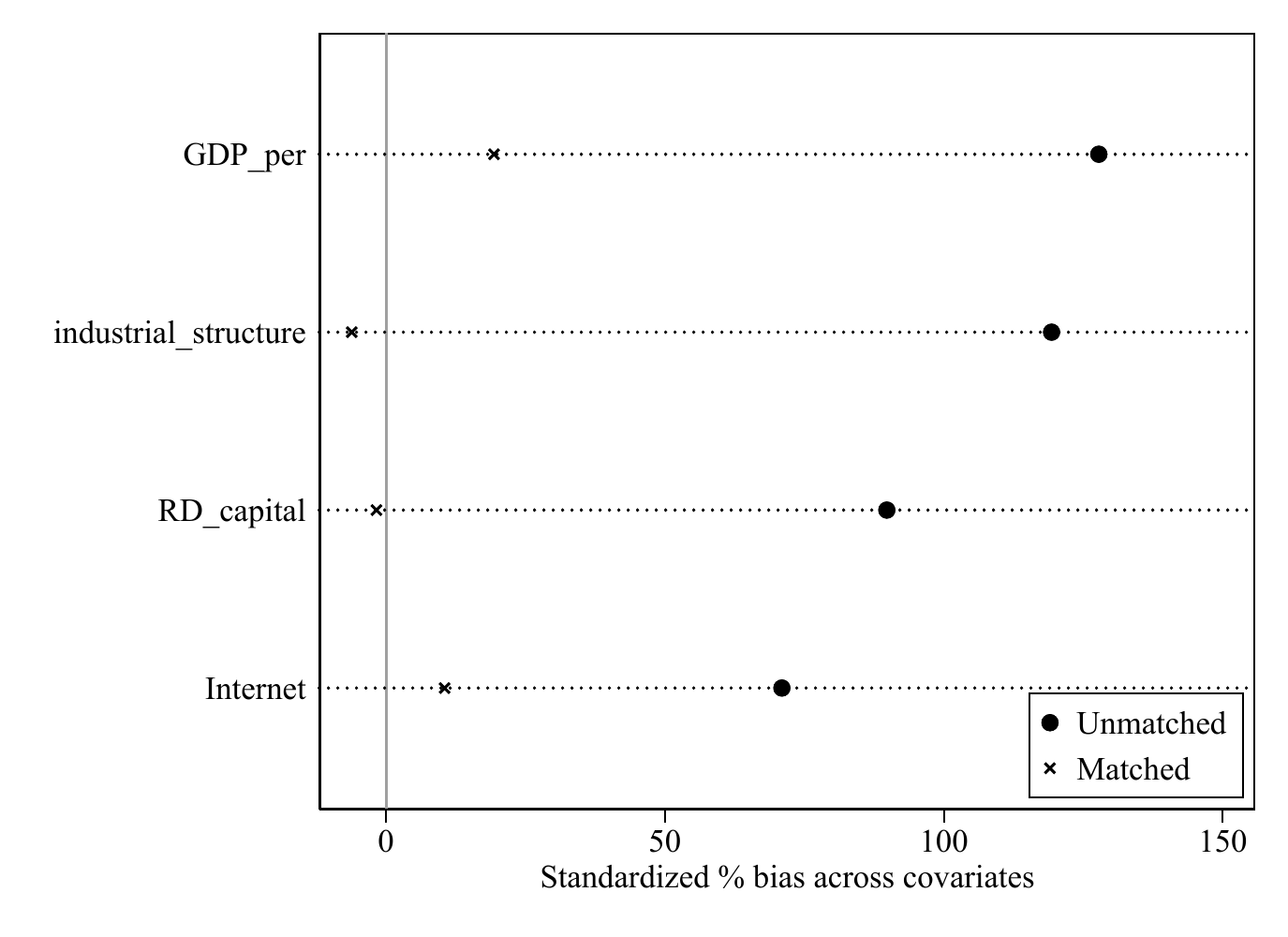}
			\caption{ PSM balance check: cross section}
		\end{minipage}%
		\begin{minipage}{.5\textwidth}
			\includegraphics[width=1\linewidth]{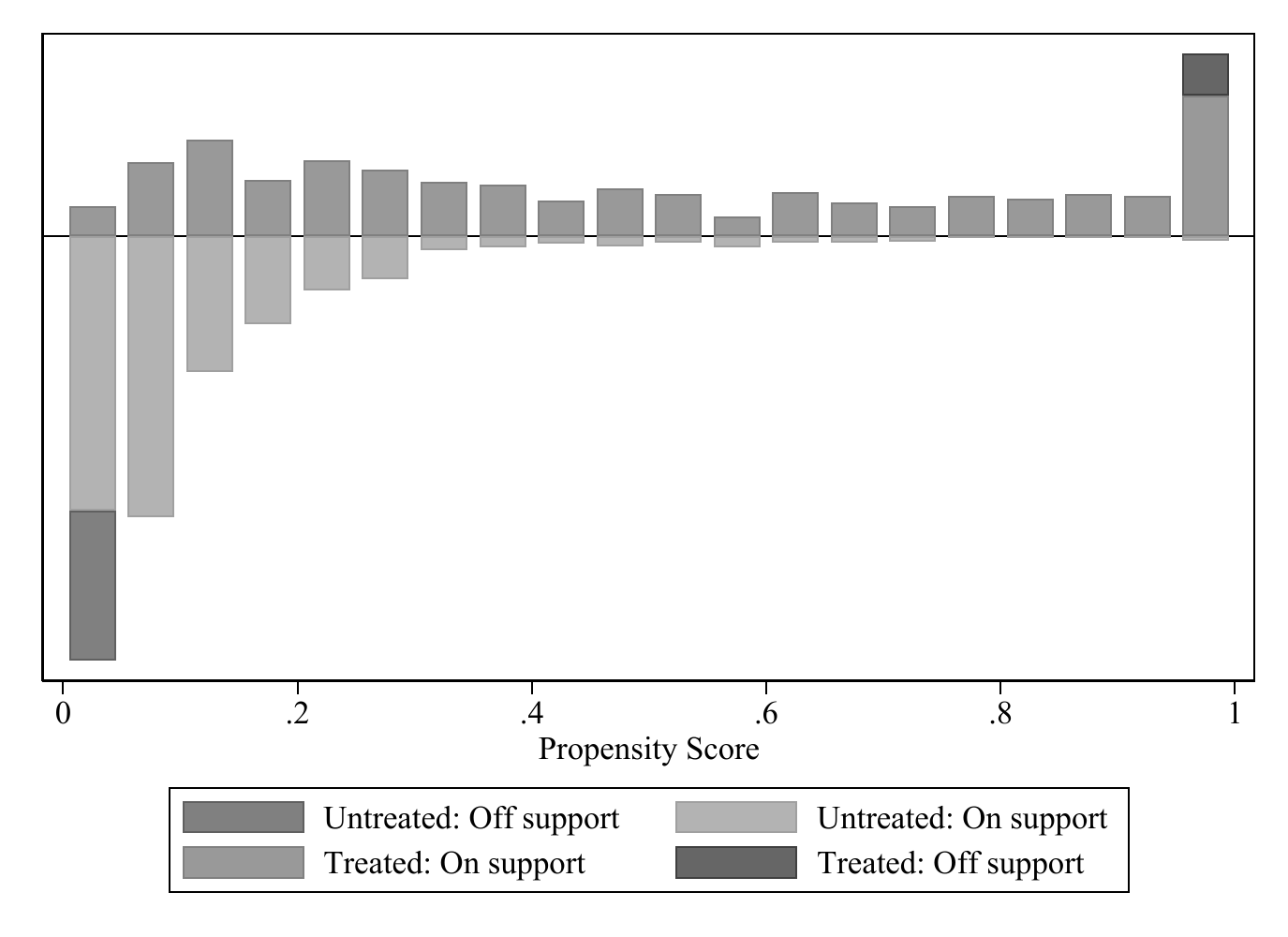}
			\caption{ Sample Layout: support the hypothesis }
		\end{minipage}
	\end{figure}
	
\end{center}

As can be seen from Figure 5, the standardized mean deviation of all matched variables after matching is less than 50\%, which is significantly smaller than the standardized deviation before matching, and there is a certain effect, but it does not yet satisfy the condition of balance test in the strict sense that the standardized mean deviation should be less than 10\%. Figure 6 shows that the vast majority of samples in the experimental and control groups are within the common range of values, while those not within the common range tend to score more extreme values, concentrated around 0. Taken together, the cross-sectional PSM has some shortcomings in terms of balance, and the results can be used only as a reference. Since the year-by-year PSM is sample-matched within each year, comparing whether each matched variable has systematic bias between the two groups can only be done in the same year.Comparing the logit regression results of different years before and after matching, i.e., if the coefficient value of each matching variable decreases, becomes insignificant and the pseudo-R2 decreases significantly after matching, it indicates that there is no systematic bias in the matching variables between the two groups in different years.

Comparing the results in Tables 8 and 9, it can be found that the coefficient values of most of the matched variables decrease in each year after matching, and most of the coefficients become insignificant, and the pseudo-R2 of all regressions significantly decreases, which to some extent indicates that there is no systematic bias in the matched variables of the two groups in different years and satisfies the requirement of the balance test.

\subsection*{Appendix IV: Propensity Matching Method Test}

\begin{center}
	\begin{figure*}[h]

	\begin{minipage}[t]{0.45\textwidth}
		\includegraphics[width=1\linewidth]{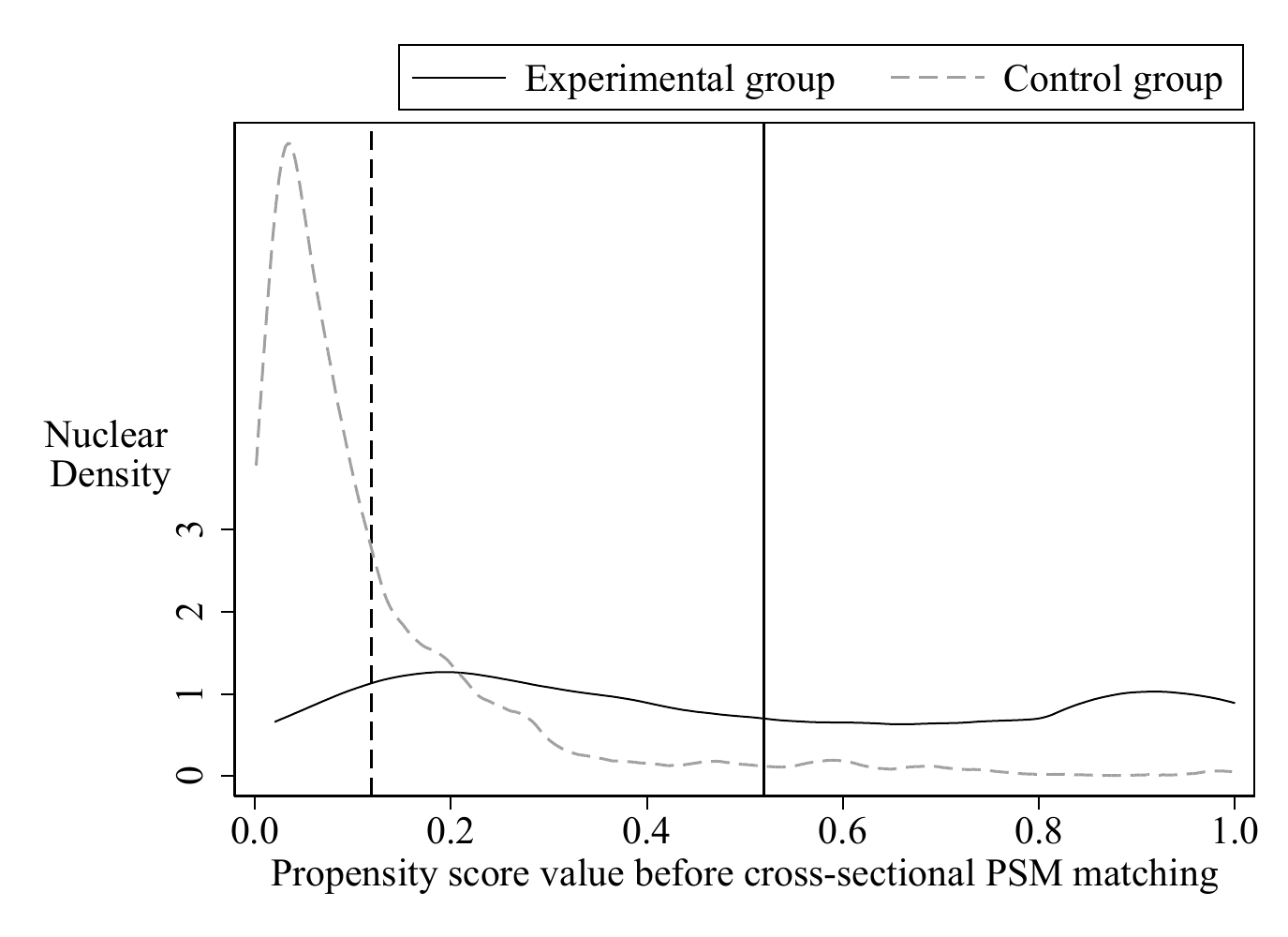}
		\caption{ before cross-sectional PSM matching}
	\end{minipage}%
	\begin{minipage}[t]{0.45\textwidth}
		\includegraphics[width=1\linewidth]{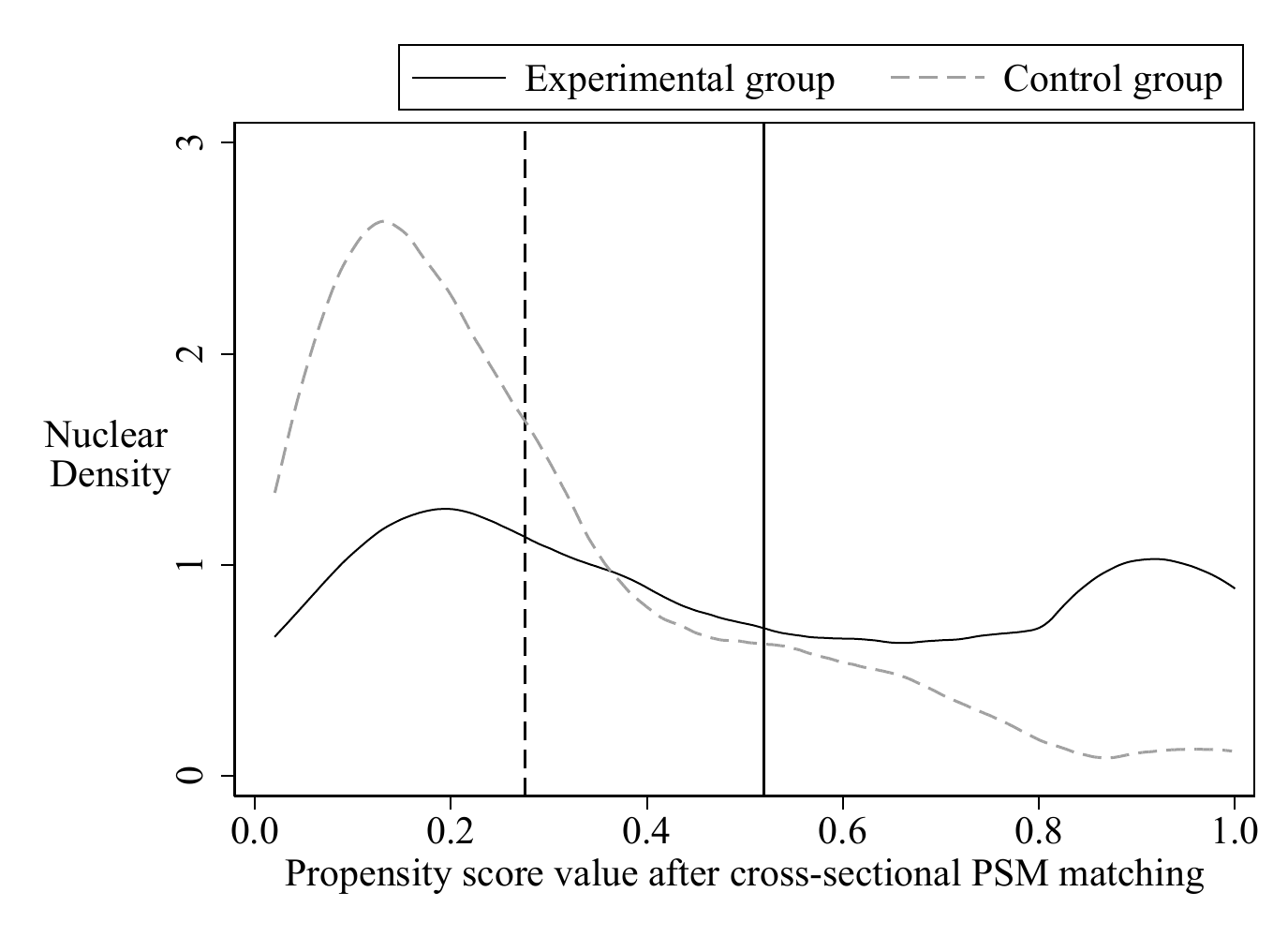}
		\caption{  after cross-sectional PSM matching}
	\end{minipage}
	\begin{minipage}[t]{0.45\textwidth}
		\includegraphics[width=1\linewidth]{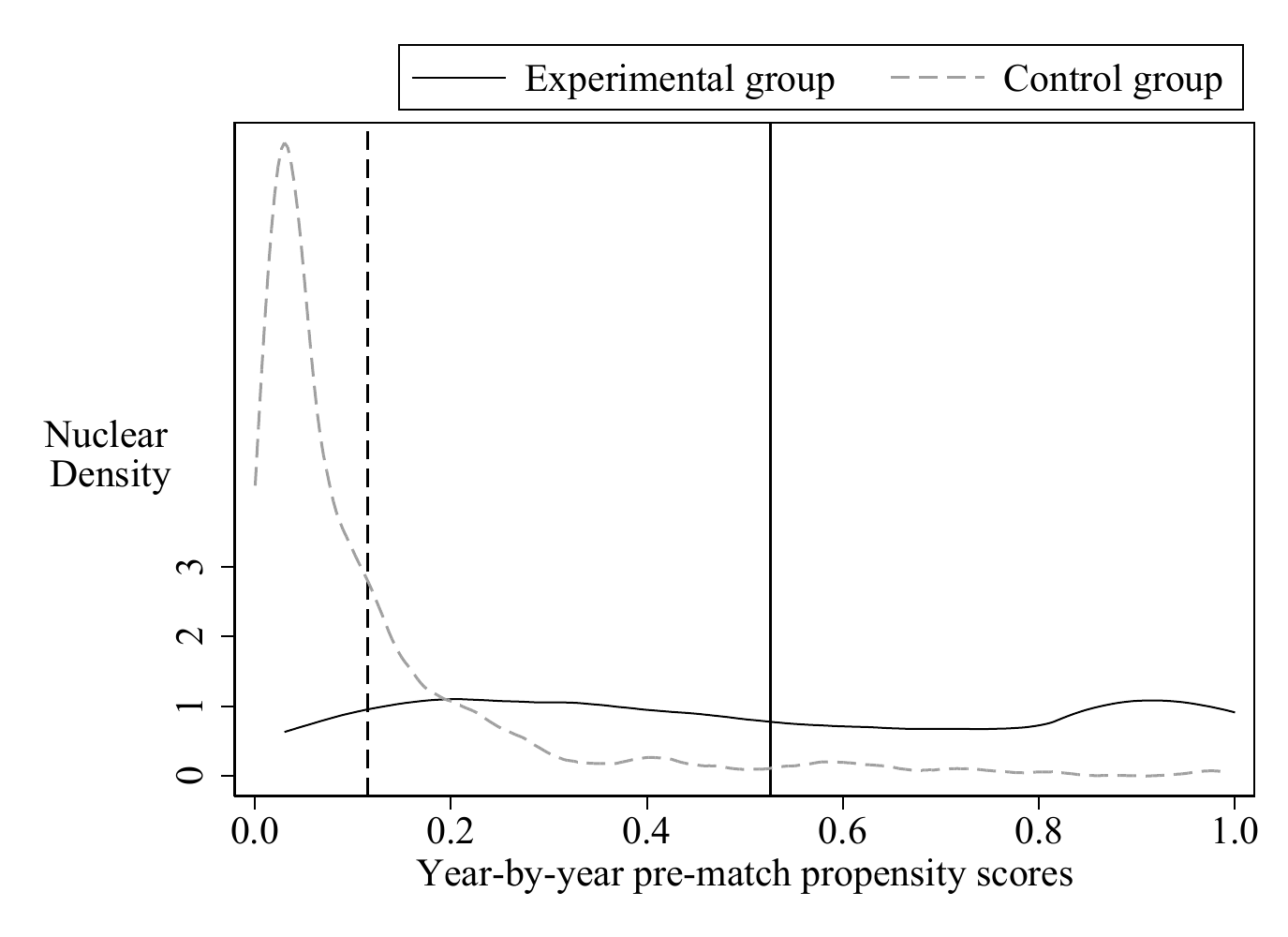}
		\caption{ before year-by-year PSM matching}
	\end{minipage}%
	\begin{minipage}[t]{0.45\textwidth}
		\includegraphics[width=1\linewidth]{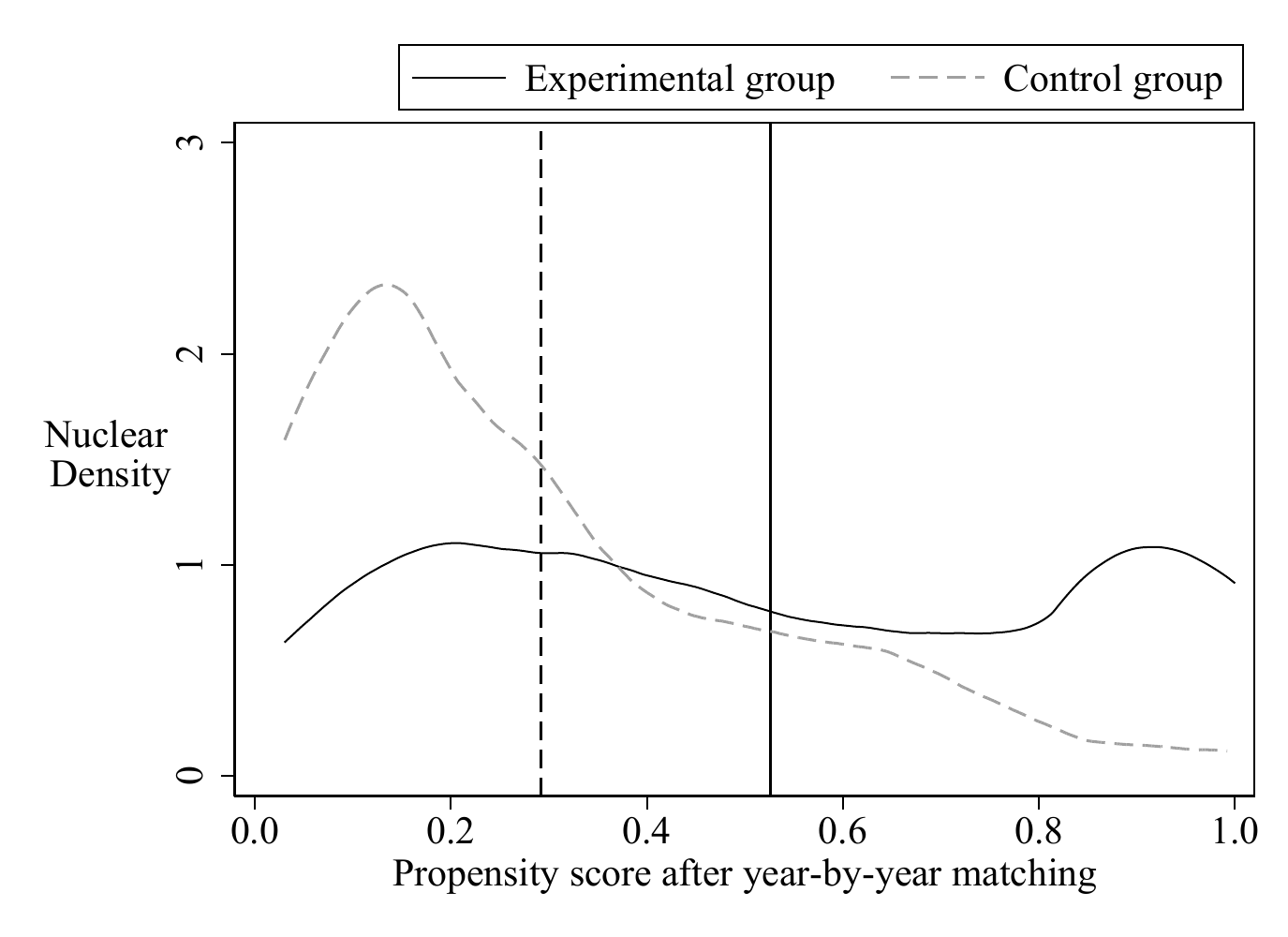}
		\caption{ after year-by-year PSM matching}
	\end{minipage}
\end{figure*}
\end{center}

Figure 7-figure supplement 10 reports the kernel density plots of the experimental and control groups before and after matching under the cross-sectional PSM and year-by-year PSM methods. It can be seen from the plots that the deviation of the two kernel density curves before and after matching is relatively large under either method, but the distance between the mean lines is shortened and the two curves are closer after matching, so it can be indicated to some extent that the cross-sectional PSM and year-by-year PSM produce the treatment effect of reducing the sample selectivity bias.

	\subsection*{Appendix V: Regression table of balance test}

\begin{landscape}
\pagestyle{landscape}
\begin{table}[h]
	\caption{Year-by-year balance test:before matching}
\begin{center}
		\begin{tabular}{ccccccccccc}
		\hline
		& (1)       & (2)       & (3)       & (4)       & (5)       & (6)       & (7)       & (8)        & (9)       & (10)      \\
		& 2011b     & 2012b     & 2013b     & 20014b    & 2015b     & 2016b     & 2017b     & 2018b      & 2019b     & 2020b     \\ \hline
		GDP\_per              & 0.5171    & 0.6320    & 0.6656    & 1.0579**  & 1.5552**  & 1.9305*** & 2.0723*** & 2.1738***  & 2.5656*** & 2.5321*** \\
		& (1.2947)  & (1.4995)  & (1.5777)  & (2.2603)  & (2.5505)  & (2.7090)  & (2.9722)  & (3.6972)   & (4.1506)  & (4.4166)  \\
		industrial\_structure & 0.0698*** & 0.0650*** & 0.0655*** & 0.0606*** & 0.0578*** & 0.0517**  & 0.0672*** & 0.0746***  & 0.0848*** & 0.0870*** \\
		& (3.5715)  & (3.5152)  & (3.2637)  & (2.9975)  & (2.7505)  & (2.4469)  & (3.0497)  & (3.2592)   & (4.0277)  & (4.1908)  \\
		Internet              & -0.2416   & -0.2796   & -2.6498   & -2.5063   & -3.9715** & -3.3674*  & -3.5890** & -3.4732*** & -3.4743** & -2.2663*  \\
		& (-0.1466) & (-0.1116) & (-1.1093) & (-1.4113) & (-2.0310) & (-1.8378) & (-2.4618) & (-2.7343)  & (-2.4591) & (-1.8117) \\
		RD\_capital           & 0.0000*** & 0.0000*** & 0.0000*** & 0.0000*** & 0.0000*** & 0.0000*** & 0.0000**  & 0.0000**   & 0.0000**  & 0.0000**  \\
		& (2.7165)  & (2.8676)  & (3.0197)  & (2.9761)  & (2.7980)  & (2.5818)  & (2.3379)  & (2.3631)   & (2.1241)  & (2.5179)  \\ \hline
		PseudoR2              & 0.3499    & 0.3564    & 0.3588    & 0.3686    & 0.3831    & 0.3942    & 0.3995    & 0.4082     & 0.4092    & 0.3889   \\ \hline
		\end{tabular}
\end{center}	
		\caption{ Year-by-year balance test:after matching}
\begin{center}
				\begin{tabular}{ccccccccccc}
			\hline
			& (1)       & (2)       & (3)       & (4)       & (5)       & (6)       & (7)       & (8)       & (9)       & (10)      \\
			& 2011a     & 2012a     & 2013a     & 20014a    & 2015a     & 2016a     & 2017a     & 2018a     & 2019a     & 2020a     \\ \hline
			GDP\_per              & -0.6172   & 0.3591    & -0.0210   & 0.1253    & 0.2232    & 0.3054    & 0.0717    & 0.5250    & 1.0246    & 1.0272    \\
			& (-1.1482) & (0.5590)  & (-0.0445) & (0.2133)  & (0.3147)  & (0.4081)  & (0.0969)  & (0.7459)  & (1.3362)  & (1.4657)  \\
			industrial\_structure & 0.0141    & -0.0141   & -0.0024   & 0.0069    & 0.0061    & 0.0162    & 0.0189    & -0.0057   & 0.0128    & 0.0084    \\
			& (0.5793)  & (-0.5337) & (-0.1049) & (0.3040)  & (0.2444)  & (0.7093)  & (0.7401)  & (-0.2022) & (0.4852)  & (0.3397)  \\
			Internet              & 1.5242    & 1.0392    & 1.1346    & -0.4182   & -1.0846   & -1.6706   & -1.2913   & -0.4505   & -1.6107   & -0.1980   \\
			& (0.8988)  & (0.4380)  & (0.4452)  & (-0.2096) & (-0.5337) & (-0.7773) & (-0.8232) & (-0.2629) & (-1.1819) & (-0.1409) \\
			RD\_capital           & 0.0000    & 0.0000    & 0.0000    & 0.0000    & 0.0000    & 0.0000    & 0.0000    & 0.0000    & 0.0000    & 0.0000    \\
			& (1.0624)  & (1.4164)  & (1.2552)  & (1.0469)  & (1.0134)  & (0.9960)  & (1.2411)  & (1.0864)  & (0.5834)  & (0.4697)  \\
			PseudoR2              & 0.0312    & 0.0529    & 0.0283    & 0.0179    & 0.0166    & 0.0253    & 0.0371    & 0.0325    & 0.0322    & 0.0341    \\ \hline
		\end{tabular}
\end{center}
		\end{table}
  \end{landscape}
\end{document}